\documentclass[draftclsnofoot, onecolumn]{IEEEtran}
\usepackage{amsmath, amsthm, amssymb} 
\usepackage{graphicx}
\usepackage{epstopdf}
\usepackage{algorithm}
\usepackage{algpseudocode}
\usepackage{multicol}

\ifCLASSOPTIONcompsoc
 \usepackage[caption=false,font=footnotesize,labelfont=sf,textfont=sf]{subfig}
\else
 \usepackage[caption=false,font=footnotesize]{subfig}
\fi

\ifCLASSOPTIONcompsoc
  \usepackage[nocompress]{cite}
\else
  \usepackage{cite}
\fi

\let\emptyset\varnothing
\allowdisplaybreaks
\begin{document}

\title{Hybrid Beamforming with Reduced Number of Phase Shifters for Massive MIMO Systems}

\author{\IEEEauthorblockN{Sohail Payami, Mir Ghoraishi$^\ast$, Mehrdad Dianati$^{\ast\ast}$, Mathini Sellathurai}\\
\IEEEauthorblockA{Heriot-Watt University, $^\ast$PureLiFi LTD, $^{\ast\ast}$University of Warwick\\
Emails: $\lbrace$s.payami, m.sellathurai$\rbrace$@hw.ac.uk, $^\ast$ mir.ghoraishi@purelifi.com, $^{\ast\ast}$m.dianati@warwick.ac.uk }}
%\markboth{Journal of \LaTeX\ Class Files,~Vol.~13, No.~9, September~2014}%
%{Shell \MakeLowercase{\textit{et al.}}: Bare Demo of IEEEtran.cls for Journals}

\maketitle


\textit{\textbf{Abstract}}\textbf{- In this paper, two novel hybrid beamforming methods are proposed to reduce the cost and power consumption of hybrid beamformers with subconnected phase shifter network structure in massive multiple-input multiple-output (MIMO) systems. This is achieved by replacing some of the phase shifters with switches which, in general, are cheaper and have lower power consumption compared to phase shifters. The proposed methods and the closed-form expressions of their performance are derived according to the properties of the elements of the singular vectors of the channel matrix. In the first approach, it is shown that by combining the subconnected phase shifter network with a fully-connected switch architecture, the number of the phase shifters can be reduced up to 50\% while the spectral efficiency is preserved. Then, in order to simplify the structure of the switch network, the fully-connected switches is replaced by subconnected switch network, e.g. binary switches. The analytical and simulation results indicate that just by using 25\% of phase shifters 90\% spectral efficiency can be achieved. Finally, simulation results indicate that similar behavior is observed when the wireless channel is considered to be sparse or correlated.}


%\vspace{-0.5cm}
\IEEEpeerreviewmaketitle
\section{Introduction}
Massive multiple-input multiple-output (MIMO) technology with digital beamforming can increase the spectral efficiency in wireless communication systems. However, a dedicated RF chain per antenna increases the cost of this technology. In order to reduce the number of RF chains, hard and soft antenna selection techniques have been proposed \cite{MOlischAntennaselection2005Journal}. In the hard selection, the RF chains are connected to the antennas by a network of switches. The drawback of this approach is that large beamforming gains cannot be achieved as only a small fraction of the antennas are used \cite{MOlischAntennaselection2005Journal,7172496}. In the soft antenna selection, also known as hybrid beamforming, the RF chains and the antennas are connected through a network of phase shifters \cite{SohailJornal,SpatiallySparsePrecodingAyach,MOlischAntennaselection2005Journal,7445130}. Such architectures have lower cost and power consumption compared to digital beamformers and they achieve a higher spectral efficiency compared to hard selection. There are two types of phase shifter networks known as fully-connected and subconnected \cite{7445130}. In the fully-connected structure, each RF chain is connected to all the antennas as in \cite{SohailJornal,MOlischAntennaselection2005Journal,SpatiallySparsePrecodingAyach}. It can exploit the full array gain, however, its power consumption can be very high due to the massive number of the phase shifters it requires \cite{SohailJornal,7445130}. In the subconnected configuration, each RF chain is connected to a subset of antennas which results in a simpler circuit but with lower spectral efficiency compared to that of the fully-connected configuration \cite{7445130}. In general, the design of the optimal soft antenna selection schemes is a challenging task due to the nonconvex constant modulus constraint imposed by the phase shifters \cite{7445130,SohailJornal,MOlischAntennaselection2005Journal,SpatiallySparsePrecodingAyach}. In this context, \cite{SohailJornal,HeathITA2015,7445130} used switches that are cheaper and power efficient alternatives to phase shifters. Similar to the phase shifter networks, switch networks have fully-connected and subconnected structures. However, due to the requirement of large number of switches, a fully-connected configuration with switches has high hardware complexity and experiences insertion losses and cross talk distortion \cite{7417765}. Hence, subconnected switch network,for example binary switches, is also preferred in practice despite the less degrees of freedom in antenna selection {\cite{7417765}. It is noted that the work in \cite{SpatiallySparsePrecodingAyach,MOlischAntennaselection2005Journal,7445130, 7417765,HeathITA2015,7172496}, and references therein, only focus on antenna selection with either phase shifters or switches, and do not consider the combination of the two methods.

Recently, the authors in \cite{HeathLetter} used for the first time a combination of switches and non-tunable phase shifters to show that the spectral efficiency of hybrid beamforming with fully-connected phase shifter network can be achieved using such combination. While the approach in \cite{HeathLetter} requires a relatively low computational complexity and low power consumption, it demands a massive number of RF routes which could result in a complex hardware and high levels of crosstalk. Within the same context of designing the joint switches and phase-shifter based hybrid beamforming, \cite{SohailJornal} considers a fully-connected phase shifter network where each phase shifter was equipped with a switch. It was shown that it is possible to reduce the power consumption of the RF beamformer by turning off almost 50\% of the phase shifters while achieving the same spectral efficiency. With the motivation to reduce the power consumption of massive MIMO systems while achieving high spectral efficiency, there is still a need for investigating new techniques to jointly design the beamforming weights at the baseband as well as the phase shifter and switch networks. 

In this paper, two such novel combinations of phase shifters and switches are proposed and their performances in terms of the achievable sum-rate are evaluated. To this end, first we derive the closed-form expression of the beamformer and its achievable sum-rate over uncorrelated independent and identically distributed (i.i.d.) channel when the RF beamformer has subconnected structure. This approach offers lower computational complexity and similar performance compared to the successive interference cancellation based method in \cite{7445130} when a small number of RF chains are connected to a large number of antennas. Second, based on the presented approach for the subconnected phase shifter network and using phase shifter selection technique, in \cite{SohailJornal}, for fully-connected phase shifter network, it is shown that a combination of subconnected phase shifter and fully-connected switch networks can reduce the number of the phase shifters by 50\% without any performance loss over uncorrelated i.i.d. Rayleigh fading channels. The simulation results for sparse scattering and correlated Rayleigh fading channels indicate that the achievable sum-rates almost remain at the same level when the number of the phase shifters are reduced to half. However, as the proposed structure requires a fully-connected switch network, which may not be suitable for practical applications, it is desirable to substitute the complex switch network with simpler structures, for example binary switches. Hence, we present another novel beamforming method that provides a slightly lower performance but with a much simpler hardware structure. In this approach, the fully-connected switches are replaced with simple subconnected switches, e.g. binary switches. Finally, the simulation results indicate that the asymptotic closed-form expressions of the spectral efficiency provide a good approximation of the performance for moderate number of antennas and phase shifters.

This paper is organized as following, the system model and hybrid beamforming with subconnected phase shifter network are described in sections \ref{sec:SystemModel} and \ref{sec:figb}. In section \ref{sec:figc}, the proposed method for hybrid beamforming with subconnected phase shifters and fully-connected switches are presented. The analysis for the subconnected phase shifters that are connected to subconnected switch network are discussed in \ref{sec:figd}. Finally, the simulation results and conclusion are presented in sections \ref{sec:SecSR} and \ref{sec:con}. 

\textit{Notations:} Bold capital and small letters \textbf{A} and \textbf{a} represent a matrix and a vector, respectively. $ A_{mn}$ denotes the $(m,n)$-th element of \textbf{A}, $\textbf{a}_m$ is the $m$-th column of \textbf{A} and $\textbf{A}_{1:m}$ is a matrix containing the first $m$ columns of \textbf{A}. det$( \textbf{A} )$, $\textbf{A}^\text{H}$ and trace(\textbf{A}) denote determinant, Hermitian and trace of \textbf{A}, respectively. Moreover, $\vert A \vert$ and $\angle A$ denote the magnitude and angle of complex number $A$. $\textbf{I}_m$ is an $m \times m$ identity matrix. Finally, $f_A(a)$, $F_A(a)$ and E$[A]$ denote the probability density function (pdf), cumulative distribution function (cdf) and expected value of $A$, respectively.

\section{System Model}
\label{sec:SystemModel}

In this work, a narrowband single-cell multiuser scenario in downlink where the base station with $N$ omni-directional antennas serves $K$ single-antenna users is considered. The wireless channel matrix $\textbf{H} \in \mathbb{C}^{K \times N}$ follows an uncorrelated Rayleigh fading model with i.i.d. elements as $H_{kn}\sim \mathcal{CN}(0,1), \: \forall k \in \lbrace 1,\: ...,\: K \rbrace$ and $\forall n \in \lbrace 1,\: ...,\: N \rbrace$. In this case, the relationship between the channel input vector $\textbf{x}\in \mathbb{C}^{N \times 1}$ and output vector $\textbf{y}\in \mathbb{C}^{K \times 1}$ is expressed as $\textbf{y}=\textbf{Hx}+\textbf{z}$ where $\textbf{z} \in \mathbb{C}^{K \times 1}$ is i.i.d. additive white Gaussian noise vector with $z_{k}\sim \mathcal{CN}(0,\sigma_\text{z}^2)$ and noise variance of $\sigma_\text{z}^2$. It is assumed that the transmitter has perfect channel state information. A vector of $K$ symbols $\textbf{u} \in \mathbb{C}^{K \times 1}$ with E$[\textbf{uu}^\text{H}]=\textbf{I}_{K}$ are precoded using the precoding matrix \textbf{F}. Then, the signal at the transmitter antennas is $\textbf{x}=\sqrt{\frac{P}{\Gamma}}\textbf{Fu}$
where $P$ is the total transmit power per stream and $\Gamma=\text{E}\big[\text{trace}  (\textbf{F}\textbf{F}^\text{H}  )\big] /K$ is a power normalization factor. Assuming that equal power is allocated to the users, the ergodic sum-capacity of downlink channel is \cite{1391204}
\begin{equation}
\label{eq:MUCap}
C(\textbf{H},\textbf{P})=\text{E}\bigg[\log_2 \text{det} \Big (\textbf{I}_K+\frac{P}{\sigma_\text{z}^2}\textbf{H}\textbf{H}^\text{H} \Big)\bigg].
\end{equation}
In massive MIMO systems, it has been shown that linear precoders such as zero-forcing (ZF) can achieve a close to optimal performance \cite{6375940}. Applying ZF precoding matrix $\textbf{H}^\text{H}(\textbf{HH}^\text{H})^{-1}$, the sum-rate becomes \cite{SohailJornal}
\begin{equation}
\label{eq:ZFRate}
R_\text{ZF}=K \log_2 \big(1+\frac{P}{\Gamma_{\text{ZF}}\sigma_z^2}\big),
\end{equation}
where $\Gamma_{\text{ZF}}=\text{E}\big[\text{trace}  \big((\textbf{HH}^\text{H})^{-1} \big )\big] /K$ is the power normalization factor for ZF precoder. When \textbf{H} follows i.i.d. Rayleigh fading model, then $\Gamma_{\text{ZF}}=1/(N-K)$ as for central complex Wishart matrices it was shown that $\text{E}[\text{trace}  \big( (\textbf{HH}^\text{H} )^{-1} \big)=K/(N-K)$ \cite{TulinoMatrix}. To maximize multiplexing gain in the high signal-to-noise ratio (SNR) regime, it is assumed that $M=K$ where $M$ is the number of the RF chains. 

Figure \ref{fullfull} presents the diagram of a fully-connected antenna selection structure where each RF chain is connected to all antennas through a network of switches or phase shifters. Depending on the performance metric, e.g. maximizing the spectral efficiency, the hard antenna selection chooses the best $M$ out of $N$ antennas using its switching network. In this paper, a matrix that includes the state of the switches, i.e. on or off which are represented by zero and one, is referred as the select matrix. The disadvantage of hard antenna selection is that large array gains cannot be achieved when $M\ll N$. In general, soft antenna selection techniques provide a better performance compared to hard selection \cite{SohailJornal,SpatiallySparsePrecodingAyach,MOlischAntennaselection2005Journal}. However, the fully-connected structure in Fig. \ref{fullfull} requires $MN$ switches or phase shifters which becomes very large in massive MIMO scenarios \cite{7445130}. This introduces high insertion losses and hardware complexity. Hence, the subconnected configuration, as shown in Fig. \ref{fig:HBFN}, is preferred in practice. The precoder matrix $\textbf{F}=\textbf{F}_{\text{sub}}\textbf{F}_{\text{B}}$ for the structure of Fig. \ref{fig:HBFN} consists of a block diagonal RF beamforming matrix $\textbf{F}_{\text{sub}} \in \mathbb{C}^{N \times M}$ and a baseband precoder $\textbf{F}_{\text{B}} \in \mathbb{C}^{M \times K}$. The RF beamformer has to be designed such that the spectral efficiency $R_\text{sub}$ is maximized subject to $ F_{\text{sub},nm}=\text{e}^{j \theta_{nm}}, \: \forall \theta_{nm} \in [0,2\pi)$ and $ \forall n \in\mathcal{I}_m$ where $\mathcal{I}_m=\lbrace \frac{N}{M}(m-1)+1,...,\frac{N}{M}m\rbrace$, otherwise $\vert F_{\text{sub},nm}\vert= 0$. In this case, $\Gamma_\text{sub}=\big[\text{trace}  (\textbf{F}_\text{sub}\textbf{F}_\text{sub}^\text{H}  )\big] /M=N/M$. In general, soft selection (hybird beamforming) is a challenging task as the maximization of the spectral efficiency is a nonconvex problem due to the constant modulus constraint imposed by the phase shifters \cite{7445130,SohailJornal,SpatiallySparsePrecodingAyach,MOlischAntennaselection2005Journal}. 
\begin{figure}
    \centering
    \subfloat[ Fully-connected network of switches/phase shifters,]
        {\includegraphics[width=.4\textwidth]{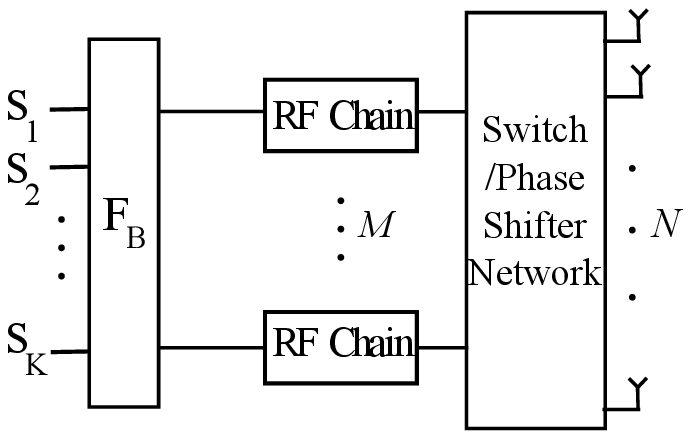}\label{fullfull}} \hfill
     \subfloat[Subconnected structure,]{
        \includegraphics[width=.4\textwidth]{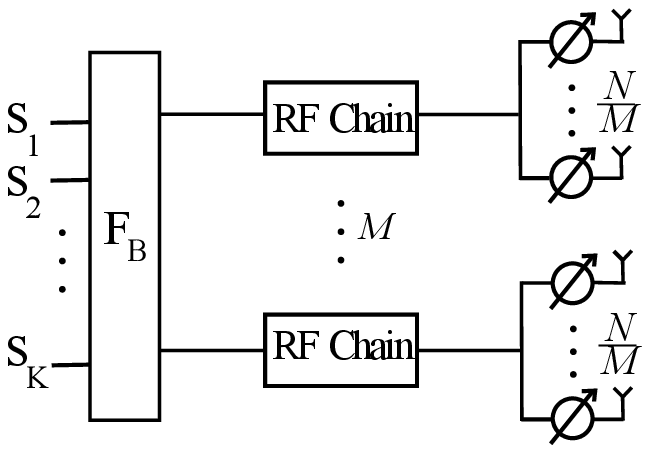} \label{fig:HBFN}}
 \\ \subfloat[ Subconnected structure with fully-connected switch network]{
        \includegraphics[width=.4\textwidth]{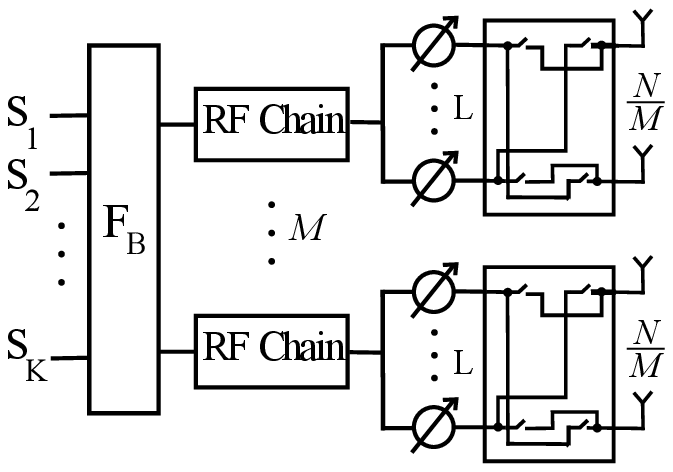}\label{fig:SubarrayWithAntennaSelection}}
    \hfill  \subfloat[ Subconnected phase shifter and switches]{
        \includegraphics[width=.4\textwidth]{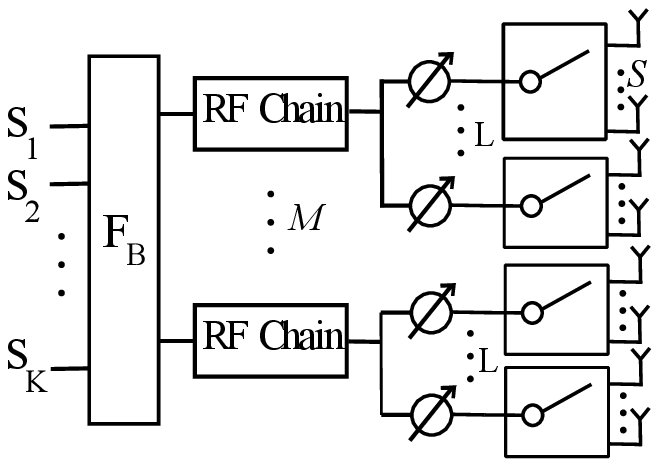}\label{fig:HybridBeamformerWithAntennaSelection}}
  \caption{Block diagram of antenna selection techniques.}
\label{fig:AllApproaches}
\end{figure}

In this paper, firstly the closed-form expression for an asymptotically optimal beamformer will be presented. In order to reduce the power consumption of the structure in Fig. \ref{fig:HBFN}, it will be shown that the configuration of Fig. \ref{fig:SubarrayWithAntennaSelection} can replace 50\% of the phase shifters with switches and without a performance loss. Finally, in Fig. \ref{fig:HybridBeamformerWithAntennaSelection}, we propose a simpler structure that the complicated fully-connected switch network is replaced with low-cost 1-out-of-$S$ switches where $S$ is the ratio of the number output-to-input ports.

\section{Subconnected Structure with Phase Shifters}
\label{sec:figb}
The singular value decomposition (SVD) of the channel matrix is denoted as $\textbf{H}=\textbf{U} \boldsymbol{\Sigma}\textbf{V}^\text{H}$, where $\textbf{V}\in \mathbb{C}^{N \times N }$ and $\textbf{U}\in \mathbb{C}^{K \times K }$ contain the right and left singular vectors. The diagonal matrix $\boldsymbol{\Sigma}\in \mathbb{R}^{K  \times N}$ includes the singular values of \textbf{H}. It is noted that $\textbf{H}=\textbf{U}\boldsymbol{\Sigma}_{1:M}\textbf{V}_{1:M}^\text{H}$ as \textbf{H} has only $M$ nonzero singular values. The statistical properties of the elements of $\textbf{V}$ when $N \to \infty$ was analyzed in \cite{SohailJornal} and it was shown that
\begin{enumerate}
   \item $\sqrt{N}V_{nn'}\sim \mathcal{CN}(0,1),\: \forall n,n' \in \lbrace 1, \: ...,\: N \rbrace$ are i.i.d.,
     \item $\vert \sqrt{N}V_{nn'}\vert $ is a Rayleigh variable with parameter $\frac{1}{\sqrt{2}}$,
\item $\sqrt{N}\text{E}[ \vert V_{nn'} \vert] = \sqrt{\pi}/2$. 
\end{enumerate}
When the impact of the beamforming matrix of the subconnected phase shifter network is considered, the achievable sum-rate in (\ref{eq:MUCap}) is expressed as 
\begin{align}
\label{eq:F}
R_\text{sub}&=\log_2 \text{det} \Big (\textbf{I}_K+ \frac{P}{\Gamma_\text{sub} \sigma_\text{z}^2}\textbf{HF}_\text{sub}\textbf{F}_\text{sub}^\text{H}\textbf{H}^\text{H} \Big) \\ \nonumber
&= \log_2 \text{det} \Big ( \textbf{I}_K+ \frac{P}{\Gamma_\text{sub} \sigma_\text{z}^2}\textbf{U}\boldsymbol{\Sigma}\textbf{V}^\text{H}\textbf{F}_\text{sub}\textbf{F}_\text{sub}^\text{H}\textbf{V}
\boldsymbol{\Sigma}^\text{H}\textbf{U}^\text{H} \Big) \\ \nonumber
&\stackrel{(a)}{=} \log_2 \text{det} \Big ( \textbf{I}_K+ \frac{P\boldsymbol{\Sigma}_{1:M}^\text{H}\boldsymbol{\Sigma}_{1:M}\textbf{V}_{1:M}^\text{H}\textbf{F}_\text{sub}\textbf{F}_\text{sub}^\text{H}\textbf{V}_{1:M}
 }{\Gamma_\text{sub} \sigma_\text{z}^2}\Big) \\ \nonumber
& \stackrel{(b)}{\leq} \log_2 \Big( \prod_{{m}=1}^{M} (1+\frac{P  \sigma_{mm}^2  Q_{mm}}{\sigma_\text{z}^2}) \Big ),
\end{align}
where (a) results from $\log_2$det$(\textbf{I}+\textbf{AB} )=\log_2$ det$(\textbf{I}+\textbf{BA} )$, $\Gamma_\text{sub}=N/M$, matrix \textbf{Q} is defined as $\textbf{Q}= \frac{M\textbf{V}_{1:M}^\text{H}\textbf{F}_\text{sub}\textbf{F}_\text{sub}^\text{H}\textbf{V}_{1:M}}{N} $, and the inequality $(b)$ comes from linear algebra as for any positive semidefinite matrix $\textbf{A} \in \mathbb{C}^{M \times M}$, $ \text{det}( \textbf{A} ) \leq \prod_{m} A_{mm}$. If $\textbf{Q}$ is a diagonal matrix, then $(b)$ in (\ref{eq:F}) turns into equality. Hence, $\textbf{F}_\text{sub}$ that upper bounds $R_\text{sub}$ should \textit{i)} diagonalize $\textbf{Q}=\textbf{GG}^\text{H}$ where $\textbf{G}=\sqrt{M/N} \textbf{V}_{1:M}^\text{H}\textbf{F}_\text{sub} $. This requires $ M/N\vert \textbf{v}_{m'}^\text{H}\textbf{f}_{\text{sub},m}\vert^2 =0$ when $m \neq m'$ and $\forall m, m' \in \lbrace 1,...,M\rbrace$, \textit{ii)} maximize the diagonal elements of $\textbf{GG}^\text{H}$. In Appendix \ref{App}, it is shown that these conditions are held and (\ref{eq:F}) is maximized when 
\begin{equation}
\label{eq:RFBFSubarray2}
F_{\text{sub},nm} =
  \begin{cases}
    \text{e}^{j\angle V_{nm}}    &  \text{if } n \in \mathcal{I}_m,  \\
    0 &  \text{if } n \notin \mathcal{I}_m.\\
  \end{cases}
\end{equation}
Moreover, according to Appendix \ref{App}, using (\ref{eq:RFBFSubarray2}) results in
\begin{align}
\label{eq:SVEV1}
\lim_{N \to \infty}\frac{M\sum_{\forall n \in \mathcal{I}_m} \vert \sqrt{N} V_{nm} \vert}{N\sqrt{M}}  =\frac{\text{E}[\sqrt{N}\vert V_{nm}\vert ]}{\sqrt{{M}}} = \frac{\sqrt{\pi}}{2\sqrt{M}},
\end{align}
due to the law of large numbers. Similarly, $\lim_{N\to \infty} \vert \sqrt{N}\textbf{v}_n^\text{H}\textbf{f}_{\text{sub},m}\vert =0, \: \forall m \neq n$ as $\text{E}[\sqrt{N}V_{mn}]=0$. 

\textit{Remark 1:} The proposed RF beamformer in (\ref{eq:RFBFSubarray2}) for Fig. \ref{fig:HBFN}, is derived under the assumption that $M$ is fixed and $L\to \infty$. Hence, it provides a suboptimal solution for relatively small values of $L$ compared to the RF beamformer of \cite{7445130}. It is noted that the computational complexity of \cite{7445130} grows with $\mathcal{O}(ML^2)$ whereas the complexity of calculating $\textbf{V}_{1:M}$ is related to $\mathcal{O}(M^3L)$ \cite{SohailJornal,Brand200620}. In our scenario of interest where $M$ is fixed and $L\to \infty$, the associated computational complexities of (\ref{eq:RFBFSubarray2}) and \cite{7445130} are proportional to $\mathcal{O}(L)$ and $\mathcal{O}(L^2)$, respectively.

In this following, the spectral efficiency for the configuration of Fig. \ref{fig:HBFN} will be calculated when (\ref{eq:RFBFSubarray2}) is used. From (\ref{eq:SVEV1}), it could be easily shown that $\sqrt{M/N}\textbf{H}\textbf{F}_\text{sub}=\sqrt{\pi}/(2\sqrt{M})\textbf{U}\boldsymbol{\Sigma}_{1:M}$. Applying ZF to $\sqrt{M/N}\textbf{HF}_\text{sub}$ to cancel the interference between the users, the precoding matrix becomes $\textbf{F}=\textbf{F}_{\text{sub}}\textbf{F}_{\text{B}}$ where $\textbf{F}_{\text{B}}=(\textbf{HF}_\text{sub})^{-1}.$ Then, the power normalization factor is
\begin{align}
\label{eq:gammasa}
\Gamma&=\text{E}\Big[\text{trace}\Big(\textbf{F}_\text{sub}(\textbf{HF}_\text{sub})^{-1}(\textbf{F}_\text{sub}^\text{H}\textbf{H}^\text{H})^{-1}\textbf{F}_\text{sub}^\text{H}\Big)\Big]/M\\ \nonumber
&=\text{E}\Big[\text{trace}\Big((\textbf{F}_\text{sub}^\text{H}\textbf{H}^\text{H})^{-1}\textbf{F}_\text{sub}^\text{H}\textbf{F}_\text{sub}(\textbf{HF}_\text{sub})^{-1}\Big)\Big]/M\\ \nonumber
&= \text{E}\Big[\text{trace}\Big((\sqrt{\frac{M}{N}}\textbf{F}_\text{sub}^\text{H}\textbf{H}^\text{H})^{-1}(\sqrt{\frac{M}{N}}\textbf{HF}_\text{sub})^{-1}\Big)\Big]/M\\ \nonumber
&=\frac{4}{\pi}\text{E}\Big[\text{trace}\Big((\textbf{H}\textbf{H}^\text{H})^{-1} \Big)\Big]=\frac{4M}{\pi} \Gamma_\text{ZF},
\end{align}
as $\lim_{N \to \infty} M/N\textbf{F}_\text{sub}^\text{H}\textbf{F}_\text{sub}=\textbf{I}_M$. Hence, the achievable sum-rate by the proposed hybrid beamformer is
\begin{align}
\label{eqZFsub}
R_\text{sub}= M \log_2 \big (  1+\frac{\pi P}{4M \Gamma_\text{ZF} \sigma_z^2}   \big ),
\end{align}
when $N\to \infty$. It is observed that in the high SNR regime, the fully-digital scheme results in $-M\log_2(\frac{\pi}{4M})$ bits/Hz/s higher spectral efficiency compared to the hybrid beamforming with subarray structure.

\textit{Remark 2:} The presented approach to derive (\ref{eqZFsub}) will be used in the rest of this paper. These steps can be summarized as 
\begin{enumerate}
	\item Diagonalization of $\textbf{Q}$.
	\item Use the i.i.d. and zero-mean properties of ${V_{nm}}$ to conclude $1/\sqrt{\Gamma_{\text{sub}}} \vert \sqrt{N}\textbf{v}_n^\text{H}\textbf{f}_{\text{sub},m}\vert \to 0, \: \forall m \neq n$.
	\item Calculate $\lim{N\to \infty}1/\sqrt{\Gamma_{\text{sub}}} \vert \sqrt{N}\textbf{v}_m^\text{H}\textbf{f}_{\text{sub},m}\vert = \text{E}[ \vert \sqrt{N} V_{nn'} \vert]$. 
	\item Calculate the power normalization factor of the hybrid beamforming, as in (\ref{eq:gammasa}), when the RF beamformer and baseband ZF precoder are combined.
	\item Replace $\Gamma_\text{ZF}$ in (\ref{eq:ZFRate}) with the power normalization factor from step 4.
\end{enumerate}
 
\section{Subconnected Phase Shifter Network - Fully-connected Switch Networks} 
\label{sec:figc}
The performance of the proposed soft selection for Fig. \ref{fig:HBFN} depends on $\vert V_{nm} \vert$. It is noted that the phase shifters that are multiplied with smaller $\vert V_{nm} \vert$ have a relatively smaller contribution to the spectral efficiency. Moreover, turning off such shifters in Fig. \ref{fig:HBFN} is equivalent to switching the corresponding antenna off. Thus, the structure of Fig. \ref{fig:SubarrayWithAntennaSelection} is proposed to reduce the number of the phase shifters by employing switch networks. By this means, the power consumption of the RF beamformer is reduced as switches require significantly smaller power to operate compared to phase shifters \cite{SohailJornal,HeathLetter,HeathITA2015}. Let $L$ denote the number of the phase shifters connected to each RF chain, and $\alpha$ be a predefined threshold. Then, by employing a fully-connected switch network and $ML$ phase shifters, the RF beamformer in (\ref{eq:RFBFSubarray2}) is modified such that the phase shifters which are corresponding to $\vert \sqrt{N}V_{nm}\vert \leq \alpha$ are turned off, i.e. $F_{\text{sub},nm}=0$, where $\alpha$ is a predefined threshold. 

Defining $V$ as an i.i.d. random variable with the same Rayleigh distribution as $\vert \sqrt{N}V_{nm}\vert$, then $f_V(  \alpha \leq v )=\text{exp}(-\alpha^2)=ML/N$ is a measure of the reduction in the number of the phase shifters. In the rest of the analysis, it is noted that $\alpha$ should be chosen carefully to make sure that $M, \: L, \: N$ are integer numbers. For the practical implementations, however, once the hardware parameters are set, the corresponding $\alpha$ will be fixed. When the number of antennas goes large and the deterministic behavior of massive MIMO are observed, $f_V(  \alpha \leq v )=\text{exp}(-\alpha^2)=ML/N$ will hold.

Let $\textbf{F}_\text{SF}\mathbb{C}^{N \times M}$ denote the RF beamforming matrix for the subconnected phase shifters with fully-connected switch networks. For $n \in\mathcal{I}_m$, the elements of $\textbf{F}_\text{SF}$ are expressed as
\begin{equation}
F_{\text{SF},nm}=
  \begin{cases}
    0   &  \text{if } \sqrt{N}  \vert V_{nm} \vert \leq \alpha,\\
    \text{exp}(j\angle V_{nm}) &  \text{if } \alpha < \sqrt{N}  \vert V_{nm} \vert .\\
  \end{cases}
\end{equation}
The received signal power is related to $\frac{1}{\sqrt{\Gamma_\text{SF}}}\textbf{v}_m^\text{H}\textbf{f}_{\text{SF},m}$ where $\Gamma_\text{SF}=L=f_V(  \alpha \leq v )N/M$. This term can be obtained as a function of $\alpha$
\begin{align}
\label{eq:gammaSF}
\frac{\textbf{v}_m^\text{H}\textbf{f}_{\text{SF},m}}{\sqrt{\Gamma_\text{SF}}}&=\lim_{N\to \infty}  \frac{M\sum_{\forall n \in \mathcal{I}_m}  \vert \sqrt{N}V_{mn}^\ast F_{\text{SF},nm} \vert }{\sqrt{Mf_V(\alpha \leq v))}N}  \\ \nonumber
&=  \frac{\text{E}[ \tilde{V} ]}{\sqrt{Mf_V(\alpha \leq v))}}  \stackrel{(a)}{=}  \frac{\frac{\sqrt{\pi}}{2}+ \alpha \text{e}^{-\alpha^2} -   \frac{\sqrt{\pi}}{2}\text{erf}(\alpha)}{\sqrt{Mf_V(\alpha \leq v))}}  ,
\end{align}
where $\tilde{V}$ is defined as
\begin{equation}
\tilde{V}=
  \begin{cases}
    0   &  \text{if } \sqrt{N}  \vert V \vert \leq \alpha,  \\
    \sqrt{N} \vert V \vert &  \text{if } \alpha < \sqrt{N} \vert V \vert,\\
  \end{cases}
\end{equation}
and (a) in (\ref{eq:gammaSF}) is a consequence of Lemma 6 in \cite{SohailJornal}. In this case, (\ref{eq:gammaSF}) results in $1/\sqrt{L}\textbf{H}\textbf{F}_\text{SF}=\text{E} [ \tilde{V} ]\textbf{U}\boldsymbol{\Sigma}_{1:M}/\sqrt{Mf_V(\alpha \leq v))}$. When the impact of ZF at the baseband is considered, the achievable rate $R_\text{SF}$ is 
\begin{equation}
\label{eqZFsubPS}
R_\text{SF}=M \log_2\bigg ( 1+\frac{ \big(\frac{\sqrt{\pi}}{2}+ \alpha \text{e}^{-\alpha^2} -   \frac{\sqrt{\pi}}{2}\text{erf}(\alpha) \big)^2}{Mf_V(\alpha \leq v)) \Gamma_\text{ZF} \sigma_z^2} P \bigg ).
\end{equation}
It is noted that (\ref{eqZFsubPS}) is a generalization of (\ref{eqZFsub}) as for $L=N/M$ (equivalently $\alpha=0$), then $R_\text{SF}=R_\text{sub}$. 

\textit{Remark 3:} The proposed $\textbf{F}_\text{SF}$ includes the effects of both subconnected phase shifter and fully-connected switch networks. In order to set the beamforming weights of the phase shifters and the select matrix of the switches in Fig. \ref{fig:SubarrayWithAntennaSelection} according to $\textbf{F}_\text{SF}$, please refer to Appendix \ref{App2}.

\section{Subconnected Phase Shifter network - Subconnected Switch Networks} 
\label{sec:figd}
As it will be discussed in the next section, the performance of hybrid selection with a fully-connected switch network and $L=N/(2M)$ phase shifters is almost equal to the subarray structure. This is equivalent to $50\%$ reduction in the number of phase shifters and significantly smaller power consumption. However, employing a fully-connected switch network requires a complex hardware with high insertion losses and crosstalk distortions. Hence, we evaluate the performance of the proposed hybrid beamformer when subconnected switch networks are employed. In this structure, as shown in Fig. \ref{fig:HybridBeamformerWithAntennaSelection}, each phase shifter is connected to only one of the $S$ adjacent antennas. In other words, the $l$th, $l \in \mathcal{I}_m$, phase shifter connected the $m$th RF chain is able to choose one of the antennas which its index is in $\mathcal{J}_q=\lbrace (q-1)S+1,...,qS \rbrace, \: \forall \mathcal{J}_q \subset \mathcal{I}_m$ where $q\in \lbrace 1,...,N/S\rbrace$. Following a similar argument as for the phase shifter selection technique, the $l$th phase shifter will be connected to the corresponding antenna element $\hat{n}$ according to $\hat{n}=\underset{n\in \mathcal{J}_q}{\arg\max} \vert V_{nm} \vert$. Let $\mathcal{N}_m=\mathcal{J}_l \cap \mathcal{I}_m$ be a set that contains $\hat{n}$, where its cardinality is $L$, and $\hat{V}$ be a random variable that has the same distribution as $\underset{n\in \mathcal{J}_l}{\max} \vert\sqrt{N} V_{nm} \vert$. 

The RF beamforming matrix $\textbf{F}_{\text{SS}}\in \mathcal{C}^{N \times M}$ for this scenario can be derived according to Algorithm 1. Since $N=MLS$ and $\Gamma_\text{SS}=L$, the received power at user side is related to 
\begin{align}
\label{eq:vftilde}
\frac{\textbf{v}_m^\text{H}\textbf{f}_\text{SS}}{\sqrt{\Gamma_\text{SS}}} &= \lim_{N\to \infty} \frac{1}{\sqrt{\Gamma_{SS} N}}  \sum_{n\in \mathcal{I}_m} \sqrt{N} V_{mn}^\ast F_{\text{SS},nm} \\ \nonumber
&=\lim_{N,L\to \infty} \frac{1}{L\sqrt{MS}}  \sum_{n\in \mathcal{N}_m } \vert \sqrt{N} V_{mn} \vert  = \frac{1}{\sqrt{MS}} \text{E} [ \hat{V}] .
\end{align}
In order to calculate $\text{E} [ \hat{V}]=\int_{-\infty}^{+\infty}f_{\hat{V}} ( \hat{v}) \hat{v} d \hat{v} $, first we calculate $F_{\hat{V}}(\hat{v})$. Since $\hat{V}$ is the maximum of $S$ i.i.d. Rayleigh distributed elements when $N \to \infty$, then
\begin{align}
F_{\hat{V}} ( \hat{v})= F_V(  v)^S   =(1-\text{e}^{- \hat{v} ^2})^S
 \end{align}
where $F_V(  v) =1-\text{e}^{- V ^2}$ as $V$ follows Rayleigh distribution. Then, the pdf of $\hat{V}$ is calculated as 
\begin{algorithm}[t]
\caption{Calculate the RF beamformer for Fig. \ref{fig:HybridBeamformerWithAntennaSelection}}
\label{test}
\begin{algorithmic}[1]
    \State $\textbf{F}_{\text{SS}}=\textbf{0}_{N\times M}$, 
    \For {$m=1:M$}
    	\State $\mathcal{N}_m=\emptyset$,
	\State $\mathcal{I}_m=\lbrace \frac{N}{M}(m-1)+1,...,\frac{N}{M}m\rbrace$,
		\For {$q=1:N/S$}
        \State $\mathcal{J}_q=\lbrace (l-1)S+1,...,lS \rbrace$,
        	\If {$\mathcal{J}_q \subset \mathcal{I}_m$}
				\State {$\hat{n}= \underset{n\in \mathcal{J}_q}{\arg\max} \vert V_{nm} \vert$},
				\State {$F_{\text{SS},\hat{n}m}=\exp{j \angle V_{\hat{n}m}}$},
				\State $\mathcal{N}_m \leftarrow \mathcal{N}_m  \cup \lbrace \hat{n} \rbrace$,
			\EndIf
		\EndFor
	\EndFor
	\State Return $\textbf{F}_\text{SS}$.
\end{algorithmic}
\end{algorithm}
\begin{align}
 f_{\hat{V}} ( \hat{v})&= \frac{d}{d\hat{v}} (1-\text{e}^{- \hat{v} ^2})^S = 2S\hat{v}(1-\text{e}^{- \hat{v} ^2})^{S-1} \text{e}^{-\hat{v}^2}  \\ \nonumber
  &\stackrel{(b)}{=} 2S\hat{v} \text{e}^{-\hat{v}^2} \sum_{s=0}^{S-1} \binom{S-1}{s} (-1)^s \text{e}^{-s\hat{v}^2} \\ \nonumber
  &= 2S\hat{v}  \sum_{s=0}^{S-1} \binom{S-1}{s} (-1)^s \text{e}^{-(s+1)\hat{v}^2},
\end{align}
where $(b)$ is the binomial expansion of $(1-\text{e}^{- \hat{v} ^2})^{S-1}$. The expected value of $\hat{V}$ is expressed as
\begin{align}
\label{eq:vtildemean}
\text{E} [\hat{V}]&= \int_{-\infty}^{+\infty}f_{\hat{V}} ( \hat{v}) \hat{v} d \hat{v} \\ \nonumber 
  &= 2S  \sum_{s=0}^{S-1} \binom{S-1}{s} (-1)^s \int_{0}^{+\infty}\hat{v}^2\text{e}^{-(s+1)\hat{v}^2} d \hat{v}\\ \nonumber 
&\stackrel{(c)}{=}  \sum_{s=0}^{S-1} \binom{S-1}{s}\frac{ (-1)^sS \sqrt{\pi}}{2(s+1)^{3/2}},
\end{align}
where $(c)$ results from $\int_{0}^{+\infty}\hat{v}^2\text{e}^{-(s+1)\hat{v}^2} d \hat{v}=\sqrt{\pi}/4(s+1)^{3/2}$ \cite{gradshteyn2007}. As a result of (\ref{eq:vftilde}) and (\ref{eq:vtildemean}),
\begin{align}
\frac{1}{\sqrt{L}}\textbf{HF}_\text{SS}&=\frac{\textbf{U}\boldsymbol{\Sigma}_{1:M}\textbf{V}_{1:M}^{\text{H}}  \textbf{F}_\text{SS}}{\sqrt{L}}\\ \nonumber
&= \sum_{s=0}^{S-1} \binom{S-1}{s}\frac{ (-1)^s\sqrt{S\pi}}{2\sqrt{M}(s+1)^{3/2}} \textbf{U}\boldsymbol{\Sigma}_{1:M}.
\end{align}
The performance of the proposed system with ZF at the baseband can be derived following the steps in (\ref{eq:gammasa}) and (\ref{eqZFsub}). In this case the spectral efficiency by the proposed scheme is 
\begin{align}
\label{eq:RSS}
R_\text{SS}= M \log_2 \bigg (  1+\frac{\big ( \sum_{s=0}^{S-1} \binom{S-1}{s}\frac{ (-1)^s}{(s+1)^{3/2}}\big)^2 PS\pi}{4M\Gamma_\text{ZF} \sigma_z^2}   \bigg ).
\end{align}
It is noted that (\ref{eq:RSS}) is a generalization of (\ref{eqZFsub}) as for $S=1$, then $R_\text{SS}=R_\text{sub}$.

\textit{Remark 4:} The proposed $\textbf{F}_\text{SS}$ includes the effects of the $ML$ phase shifters and the $ML$ subconnected switches. In order to set the weights of the phase shifters and switches in Fig. \ref{fig:HybridBeamformerWithAntennaSelection} according to $\textbf{F}_\text{SS}$, please refer to Appendix \ref{App3}.

\section{Simulation Results}
\label{sec:SecSR}

In this section, computer simulations are used to evaluate the performance of the proposed antenna selection techniques for the structures in Fig. \ref{fig:HBFN} to Fig. \ref{fig:HybridBeamformerWithAntennaSelection}. In addition, the closed-form expressions in (\ref{eqZFsub}), (\ref{eqZFsubPS}) and (\ref{eq:RSS}) will be examined when $N \to \infty$ does not hold. Monte-Carlo simulations over 1000 realizations for $M=K=4$ and $P/\sigma_\text{z}^2=10 $ dB are used to assess the performance. At the end of this section, the performance of the proposed methods over sparse scattering channels as well as correlated Rayleigh fading channels will be examined. 

Figure \ref{FigImpactofSwitch} shows the tradeoffs between the spectral efficiency and the total number of the phase shifters $ML$ when $N$ is fixed. In order to guarantee that the properties of massive MIMO are observed and the hybrid beamformer of (\ref{eq:RFBFSubarray2}) is close to optimal, $N$ is set to a large number as $N=512$. It is noted that the fully-connected switch network provides more flexibility between the number of the input and output ports which is not possible with 1-out-of-$S$ switches. When $ML/N=0.75$, Fig. \ref{FigImpactofSwitch} indicates that the fully-connected switch networks with phase shifter selection provides slightly higher spectral efficiency compared to the structure of Fig. \ref{fig:HBFN}. In addition, compared to the scenario that each antenna has a phase shifter, the number of the phase shifters can be reduced to 50\% without a performance loss when a fully-connected switch networks with $ML/N=0.5$ is used. Figure \ref{FigImpactofSwitch} also shows that when a simple binary switch is used, i.e. $S=2$, the loss of the achievable rate is less than 1 bits/Hz/s compared to soft selection with subconnected structure. It is observed that the proposed method with $S=4$, or equivalently $ML=128$ phase shifters, achieves around 93\% of the spectral efficiency compared to the scenario that $ML=512$. Figure \ref{FigImpactofSwitch} also shows that there is good match between the simulation results and the closed-form expressions of (\ref{eqZFsub}), (\ref{eqZFsubPS}) and (\ref{eq:RSS}) for various ratios of the number of inputs to outputs. 
\begin{figure}[!tbp]
  \centering
    \includegraphics[width=.48\textwidth]{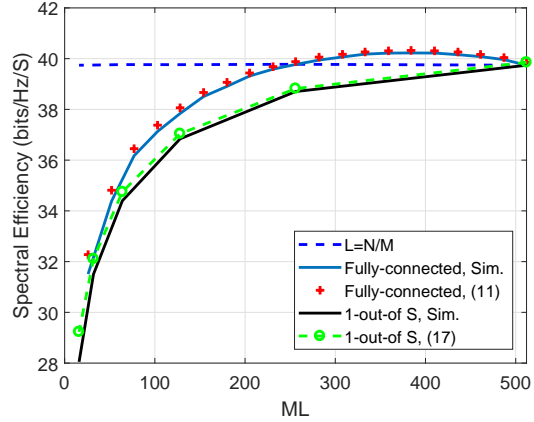}
    \caption{Spectral efficiency by the proposed techniques versus the number of the phase shifters, $N=512$, $M=4$ and $P/\sigma_\text{z}^2=10$ dB.}
        \label{FigImpactofSwitch}
\end{figure}
Figure \ref{fig:NumAntennas} shows the impact of the number of the antennas on the accuracy of the closed-form expressions of spectral efficiency. It is assumed that the ratio of inputs to outputs is $ML/N=0.5$. At $N=32$, 12\% error between the simulation results and (\ref{eqZFsubPS}) and (\ref{eq:RSS}) is observed. This is due to the fact that $L=N/2M=4$ is small and, hence, the law of large numbers does not hold. When $L$ increases to 8, the error between the simulations and analytical results reaches to around 3\%. Figure \ref{fig:NumAntennas} indicates that equations (\ref{eqZFsub}), (\ref{eqZFsubPS}) and (\ref{eq:RSS}) can provide a good approximation of the performance when $16\leq L$. 

Figure \ref{fig:SNR} presents the achievable rates by the proposed beamformer with binary switches when $P/\sigma_\text{z}^2$ varies. Compared to the structure of Fig. \ref{fig:HBFN} with $L=N/M$ phase shifters, it is observed that the performance loss due to the use of binary switches is almost negligible at the high SNR regime. In addition, Fig. \ref{fig:SNR} provides a comparison between our RF beamformer in (\ref{eq:RFBFSubarray2}) for the structure in Fig. \ref{fig:HBFN} and its asymptotic performance expression (\ref{eqZFsub}), which was derived under asymptotically large number of antennas, and the RF beamformer of \cite{7445130}. In both cases, ZF is applied to the effective channel matrix $\textbf{H}_\text{e}=1/\sqrt{\Gamma_\text{sub}}\textbf{HF}_\text{sub}$ at the baseband. It is observed that \cite{7445130} achieves a constant 1.5 bits/Hz/s higher spectral efficiency, which is at the cost of higher complexity, compared to the beamformer in (\ref{eq:RFBFSubarray2}). When a simple binary switch is used and the number of the phase shifters is reduced from 128 to 64, the achievable sum-rate is around 2.7 bits/Hz/s less than \cite{7445130} with 128 phase shifters.  

\begin{figure}
\centering
     \includegraphics[width=.48\textwidth]{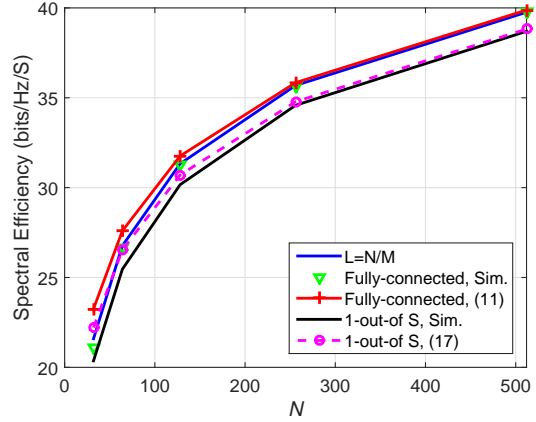}
      \caption{Spectral efficiency by the proposed techniques versus the number of the antennas, $ML/N=0.5$, $M=4$ and $P/\sigma_\text{z}^2=10$ dB.}
        \label{fig:NumAntennas}
\end{figure}
\begin{figure}
\centering
    \includegraphics[width=.48\textwidth]{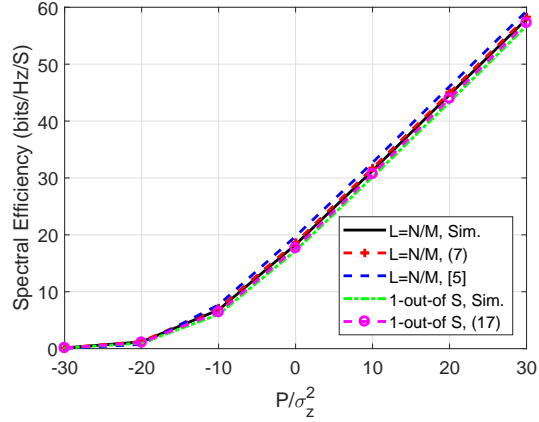}
        \caption{Spectral efficiency by the proposed techniques versus $P/\sigma_\text{z}^2$ for $L=N/M$ and $L=N/2M$, $M=4$ and $N=128$.}
        \label{fig:SNR}
\end{figure}
Although i.i.d. Rayleigh fading channel model is commonly used in the literature on massive MIMO to present theoretical studies, as in \cite{marzettabook2016} and references therein, this channel model may not be suitable for practical applications. In order to further investigate the performance of our proposed methods under more realistic channels, in the following we present the achievable sum-rates by these techniques over correlated Rayleigh fading and sparse geometry-based channel models. An exponentially correlated MIMO channel model will be used to model correlation \cite{6777295}. We assume that the correlation effects are observed at the base station according to $\textbf{H}=\textbf{H}_\text{w}\textbf{R}^{1/2}$ where $\textbf{H}_\text{w}\in \mathbb{C}^{K \times N}$ is zero-mean i.i.d. Rayleigh fading channel matrix, and $\textbf{R}\in \mathbb{C}^{N \times N}$ is the correlation matrix as
\begin{align}
\label{CorrelationMatrix}
\textbf{R}&=  \begin{pmatrix}
  1 & \rho & \cdots & \rho^{N-1} \\
  \rho & 1 & \cdots & \rho^{N-2} \\
  \vdots  & \vdots  & \ddots & \vdots  \\
\rho^{N-1} & \rho^{N-2}& \cdots &1
 \end{pmatrix}, 
\end{align}
where $0\leq \rho \leq 1$ is the correlation coefficient. To evaluate the performance of the proposed methods over sparse channels, we use the geometry-based model with uniform and linearly spaced antennas at the base station. Assuming there are $C$ multipath components (MPCs) in the channel between the base station and user $k$, the channel vector $\textbf{h}^\text{T}_k \in \mathbb{C}^{N \times 1}$ for user $k$ is
\begin{equation}
\label{eq:SparseChannel}
\textbf{h}^\text{T}_k=\sqrt{\dfrac{N }{C}}\sum_{c=1}^C  \beta_{ck}\textbf{a}^\ast(\phi_{ck}),
\end{equation}
where $\beta_{ck}\sim \mathcal{CN}(0,1)$ is the multipath coefficient, $\phi_{ck}$ is the angle-of-departure of the $c$th multipath. The steering vector $\textbf{a}(\phi_{ck})$ for linear arrays is expressed as
\begin{align}
\label{Eq:Manifoldvector}
\textbf{a}(\phi_{ck})=\dfrac{1}{\sqrt{N}}(1, \text{e}^{\frac{j2\pi d}{\lambda} \cos(\phi_{ck})}\: ...,\: \text{e}^{\frac{j2\pi d}{\lambda}(N-1)\cos(\phi_{ck})})^\text{T}
\end{align}
where $\phi_{ck}\in [0,\: \pi]$, $\lambda$ is the wavelength and $d$ is the antenna spacing. In our simulations, it is assumed that $d=\lambda/2$. 

Similar to Fig. \ref{FigImpactofSwitch}, the tradeoffs between the spectral efficiency and the total number of the phase shifters $ML$ is studied in Fig. \ref{fig:ChannelCompare} when different channel models are considered. In order to make a comprehensive comparison in a single plot, $P/\sigma_\text{z}^2$ is set to 0 dB, 10 dB, and 20 dB to perform the computer simulations to evaluate the performance over uncorrelated Rayleigh fading, correlated Rayleigh fading and sparse channels, respectively. For the sparse channel model, we assume that there are only 2 MPCs, i.e. $C=2$ in (\ref{eq:SparseChannel}), from the base station to each user. For the correlated Rayleigh fading channel, it is assumed that $\rho=0.7$ in (\ref{CorrelationMatrix}). In addition, we use the RF beamformer of \cite{7445130} as reference to evaluate the performance of our methods. Figure \ref{fig:ChannelCompare} shows that the RF beamformer in (\ref{eq:RFBFSubarray2}) and \cite{7445130} have almost the same performance for the sparse channels, and the difference in sum-rate is less than 1.2 bits/Hz/s for the other two channels. When the structure in Fig. \ref{fig:SubarrayWithAntennaSelection} is used and the number of the phase shifters is reduced by 50\%, the performance of our methods is, at most, 1.3 bits/Hz/s lower than \cite{7445130} which requires 512 phase shifters. For 75\% reduction, the beamformer of \cite{7445130}, with 512 phase shifters, has 4.65 bits/Hz/s, 3.03 bits/Hz/s and 2.55 bits/Hz/s higher achievable rate compared to the sparse, correlated and uncorrelated channels, respectively. Replacing the fully-connected with binary switches, the performance difference between our method and \cite{7445130} becomes 2.43 bits/Hz/s, 2.86 bits/Hz/s and 1.73 bits/Hz/s for the sparse, correlated and uncorrelated channels, respectively. 

\begin{figure}
\centering
    \includegraphics[width=.48\textwidth]{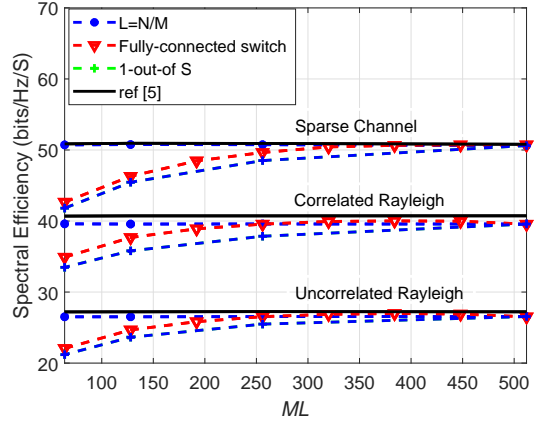}
        \caption{Spectral efficiency by the proposed techniques versus the number of the phase shifters, $N=512$, $M=4$. The parameters for the sparse channel are $P/\sigma_\text{z}^2=20$ dB and $C=2$, for the correlated Rayleigh fading $r=0.7$ and $P/\sigma_\text{z}^2=10$ dB, for the uncorrelated Rayleigh fading $P/\sigma_\text{z}^2= 0$ dB.}
        \label{fig:ChannelCompare}
\end{figure}

\section{Conclusion}
\label{sec:con}
In this paper, we investigated the performance of hybrid beamformers when the RF beamformer consists of a combination of subconnected phase shifter network with fully-connected/subconnected switch networks. The proposed beamforming methods and the closed-form expressions of their spectral efficiencies were derived based on the properties of the singular vectors of the channel matrix when the propagation environment is modeled by Rayleigh fading. The simulation results indicated that the proposed methods can perform well when channel sparsity and correlation effects are considered. Such structures reduce the power consumption of hybrid beamformers with phase shifters only as switches require significantly lower power to operate compared to the phase shifters. Specially, in massive MIMO systems where the number of the required phase shifters is large. It was shown that the fully-connected switch network provides slightly better performance compared to the subbonnected structure. However, due to the simplicity of the subconnected approach and lower insertion losses and crosstalks, it is preferred in practice. 

The power consumption of switches is negligible compared to phase shifters, as a result, our methods can significantly reduce the power consumption. On the other hand, such structures can complicate the channel estimation. Hence, in future we are aiming to analyze the joint optimization of spectral and energy efficiencies to choose the system parameters, i.e. number of the switches, phase shifters and antennas. In addition, further research is required to investigate the impact of channel estimation when the proposed structures are used.
\appendices

\section{Diagonalization of \textbf{Q} in (\ref{eq:F})}
\label{App}
In order to analyze $\textbf{Q}$, we investigate the behavior of the elements of $\textbf{G} \in \mathbb{C}^{M\times M}$, defined as $\textbf{G}=\sqrt{M/N} \textbf{V}_{1:M}^\text{H}\textbf{F}_\text{sub} $. It is noted that our choice of $\textbf{f}_{\text{sub},m}$ will result in either \text{case 1:} $\sqrt{\frac{M}{N}}\textbf{v}_{m}^\text{H}\textbf{f}_{\text{sub},m'}=0$, or \text{case 2:} $\sqrt{\frac{M}{N}}\textbf{v}_{m}^\text{H}\textbf{f}_{\text{sub},m'}\neq 0$, $\forall \: m\neq m'$. In the first case that $\sqrt{\frac{M}{N}}\textbf{v}_{m}^\text{H}\textbf{f}_{\text{RF},m'}=0$, $\forall \: m\neq m'$, it could be easily shown that all of the elements of $\textbf{G}$ except the $G_{mm}$ become zero, and (b) in (\ref{eq:F}) turns into equality. Then,
  \begin{align}
\label{eq:SNRMeasureSubArray}
G_{mm}&=\frac{\sqrt{M}\textbf{v}_m^\text{H}\textbf{f}_{\text{sub},m}}{\sqrt{N}}=\sqrt{\frac{M}{N}}\sum_{n\in \mathcal{I}_m} V_{nm}^\ast \text{e}^{j\theta_{nm}} \\ \nonumber
&\leq \sqrt{\frac{M}{N}}\sum_{n\in \mathcal{I}_m}\vert V_{nm} \vert ,
\end{align}
where the equality holds when 
   \begin{equation}
\label{eq:AppRFBFSubarray}
F_{\text{sub},nm} =
  \begin{cases}
    \text{e}^{j\angle V_{nm}}    &  \text{if } n \in \mathcal{I}_m,  \\
    0 &  \text{if } n \notin \mathcal{I}_m.\\
  \end{cases}
\end{equation}
Using this beamforming matrix and in the limit of large numbers, when $N \to \infty$, (\ref{eq:SNRMeasureSubArray}) becomes
\begin{align}
\label{eq:APPSVEV}
\lim_{N \to \infty}\frac{M\sum_{\forall n \in \mathcal{I}_m} \vert \sqrt{N} V_{nm} \vert}{N\sqrt{M}}  =\frac{\text{E}[\sqrt{N}\vert V_{nm}\vert ]}{\sqrt{{M}}} = \frac{\sqrt{\pi}}{2\sqrt{M}}.
\end{align}
In the following, we analyze the impact of setting $\textbf{F}_\text{sub}$, according to (\ref{eq:AppRFBFSubarray}), on the off-diagonal elements of $\textbf{G}$. For the uncorrelated i.i.d. Rayleigh channel, the elements of singular vectors of the channel matrix are zero-mean i.i.d. random variables and their phases are uniformly distributed over $[0,\: 2\pi]$ \cite{SohailJornal}. As a consequence of law of large numbers
\begin{align}
\label{eq:OffDiagRayleigh}
\lim_{N \to \infty}    \sqrt{\frac{M}{N}} \textbf{v}_{m}^\text{H}\textbf{f}_{\text{sub},m'} &=\lim_{N \to \infty}  \frac{\sqrt{M}}{N} \sum_{\forall n \in \mathcal{I}_{m'}} \sqrt{N}  V_{nm}^\ast \text{e}^{j \angle  V_{nm'}}  \\ \nonumber
&=\sqrt{M} \text{E}[ \sqrt{N} V_{nm}]=0.
\end{align}
As a result, it could be concluded that all of the elements of \textbf{G} except the diagonal elements become zero. This means that choosing $\textbf{F}_\text{sub}$ according to (\ref{eq:AppRFBFSubarray}) will fulfill the condition of \textit{case 1}. Moreover, this matrix can diagonalize $\textbf{Q}$ and maximize its diagonal element. Hence, the achievable sum-rate in  (\ref{eq:F}) is maximized.

  \hfill\(\Box\)

\section{The Weights for the Switches and Phase Shifters of Figure \ref{fig:SubarrayWithAntennaSelection} }
\label{App2}
In order to set the phases of the phase shifters and the select network of the switches in Fig. \ref{fig:SubarrayWithAntennaSelection} according to $\textbf{F}_\text{SF}$, the following procedure can be applied:
\begin{enumerate}
\item Select the elements of $F_{\text{SF},nm}$ $\forall n \in \mathcal{I}_m$. With the same index order, store them in vector $\tilde{\textbf{f}}_{\text{SF}}^{(m)}\in \mathbb{C}^{\frac{N}{M} \times 1}$. It is noted that $\tilde{\textbf{f}}_{\text{SF}}^{(m)}$ has only $L$ nonzero elements.
\item Create the vector $\bar{\textbf{f}}_{\text{SF}}^{(m)}\in \mathbb{C}^{L \times 1}$ which includes all the nonzero elements of $\tilde{\textbf{f}}_{\text{SF}}^{(m)}$ with the same order. The vector $\bar{\textbf{f}}_{\text{SF}}^{(m)}$ contains the beamforming weights of the phase shifters that are connected the $m$th RF chain.
\item Let the matrix $\textbf{S}^{(m)}\in \mathbb{C}^{\frac{N}{M} \times L}$ represent the select network of the fully-connected switch on the $m$th RF chain where $\tilde{\textbf{f}}_{\text{SF}}^{(m)}=\textbf{S}^{(m)} \bar{\textbf{f}}_{\text{SF}}^{(m)}$. Initially, set all the elements of $\textbf{S}^{(m)}$ to zero.
\item If $\tilde{f}_{\text{SF},i}^{(m)}\neq 0$, then set $S_{il}^{(m)}=1$ such that $\tilde{f}_{\text{SF},i}^{(m)}=\bar{f}_{\text{SF},l}^{(m)}$, $\forall i \in\lbrace1,...,N/M \rbrace$ and $\forall l \in\lbrace1,...,L \rbrace$. 
\end{enumerate}
It is noted that each row of $\textbf{S}^{(m)}$ can have at most one nonzero element. In addition, the presented method to set the phase shifter and switches is just one solution, and it is not unique.
  \hfill\(\Box\)

\section{The Weights for the Switches and Phase Shifters of Figure \ref{fig:HybridBeamformerWithAntennaSelection} }
\label{App3}
In order to set the phase of the $l$th phase shifter on the $m$th RF chain, and the select network of the switches in Fig. \ref{fig:HybridBeamformerWithAntennaSelection} according to $\textbf{F}_\text{SS}$, the following procedure can be applied:
\begin{enumerate}
\item Select the elements of $F_{\text{SS},nm}$ where $n\in \lbrace (m-1)LS+(l-1)S+1,...,(m-1)LS+lS \rbrace$ and $l \in \lbrace1,...,L\rbrace$. With the same index order, store them into $\tilde{\textbf{f}}_{\text{SS}}^{(ml)}\in \mathbb{C}^{S \times 1}$. Let $\hat{s}$ denote the index of the only nonzero element of $\tilde{\textbf{f}}_{\text{SS}}^{(ml)}$ where $\hat{s}\in \lbrace1,...,S\rbrace$.
\item Set the $l$th phase shifter on the $m$th RF chain according to the nonzero element of $\tilde{\textbf{f}}_{\text{SS}}^{(ml)}$.
\item Let vector $\textbf{s}^{(ml)}\in \mathbb{C}^{S \times 1}$, represent the select network of the subconnected switch which connects the $l$th phase shifter on the $m$th RF chain to one of the $n\in \lbrace (m-1)LS+(l-1)S+1,...,(m-1)LS+lS \rbrace$ antennas. Except $s_{\hat{s}}^{(ml)}=1$, set all the elements of $\textbf{s}^{(ml)}$ to zero.
\end{enumerate}

  \hfill\(\Box\)

\section*{Acknowledgment}
The research leading to these results has received funding from the European Union Seventh Framework Programme (FP7/2007-2013) under grant agreement $\text{n}^\circ$619563 (MiWaveS). We would also like to acknowledge the support of the University of Surrey 5GIC members for this work.

\ifCLASSOPTIONcaptionsoff
  \newpage
\fi

\bibliographystyle{IEEEtran}
\bibliography{IEEEabrv,References}

\begin{thebibliography}{10}
\providecommand{\url}[1]{#1}
\csname url@samestyle\endcsname
\providecommand{\newblock}{\relax}
\providecommand{\bibinfo}[2]{#2}
\providecommand{\BIBentrySTDinterwordspacing}{\spaceskip=0pt\relax}
\providecommand{\BIBentryALTinterwordstretchfactor}{4}
\providecommand{\BIBentryALTinterwordspacing}{\spaceskip=\fontdimen2\font plus
\BIBentryALTinterwordstretchfactor\fontdimen3\font minus
  \fontdimen4\font\relax}
\providecommand{\BIBforeignlanguage}[2]{{%
\expandafter\ifx\csname l@#1\endcsname\relax
\typeout{** WARNING: IEEEtran.bst: No hyphenation pattern has been}%
\typeout{** loaded for the language `#1'. Using the pattern for}%
\typeout{** the default language instead.}%
\else
\language=\csname l@#1\endcsname
\fi
#2}}
\providecommand{\BIBdecl}{\relax}
\BIBdecl

\bibitem{MOlischAntennaselection2005Journal}
X.~Zhang, A.~Molisch, and S.-Y. Kung, ``Variable-phase-shift-based
  {RF}-baseband codesign for {MIMO} antenna selection,'' \emph{IEEE
  Transactions on Signal Processing}, vol.~53, no.~11, pp. 4091--4103, November
  2005.

\bibitem{7172496}
X.~Gao, O.~Edfors, F.~Tufvesson, and E.~G. Larsson, ``Massive {MIMO} in real
  propagation environments: Do all antennas contribute equally?'' \emph{IEEE
  Transactions on Communications}, vol.~63, no.~11, pp. 3917--3928, November
  2015.

\bibitem{SohailJornal}
S.~Payami, M.~Ghoraishi, and M.~Dianati, ``Hybrid beamforming for large antenna
  arrays with phase shifter selection,'' \emph{IEEE Transactions on Wireless
  Communications}, vol.~PP, no.~99, pp. 1--1, 2016.

\bibitem{SpatiallySparsePrecodingAyach}
O.~El~Ayach, S.~Rajagopal, S.~Abu-Surra, Z.~Pi, and R.~Heath, ``Spatially
  sparse precoding in millimeter wave {MIMO} systems,'' \emph{IEEE Transactions
  on Wireless Communications}, vol.~13, no.~3, pp. 1499--1513, March 2014.

\bibitem{7445130}
X.~Gao, L.~Dai, S.~Han, C.~L. I, and R.~W. Heath, ``Energy-efficient hybrid
  analog and digital precoding for {M}m{W}ave {MIMO} systems with large antenna
  arrays,'' \emph{IEEE Journal on Selected Areas in Communications}, vol.~34,
  no.~4, pp. 998--1009, April 2016.

\bibitem{HeathITA2015}
R.~Mendez-Rial, C.~Rusu, A.~Alkhateeb, N.~Gonzalez-Prelcic, and R.~W. Heath,
  ``Channel estimation and hybrid combining for mmwave: Phase shifters or
  switches?'' \emph{Information Theory and Applications Workshop (ITA), 2015},
  pp. 90--97, February 2015.

\bibitem{7417765}
X.~Gao, O.~Edfors, F.~Tufvesson, and E.~G. Larsson, ``Multi-switch for antenna
  selection in massive {MIMO},'' \emph{2015 IEEE Global Communications
  Conference (GLOBECOM)}, pp. 1--6, Dec 2015.

\bibitem{HeathLetter}
A.~Alkhateeb, Y.~H. Nam, J.~Zhang, and R.~W. Heath, ``Massive {MIMO} combining
  with switches,'' \emph{IEEE Wireless Communications Letters}, vol.~5, no.~3,
  pp. 232--235, June 2016.

\bibitem{1391204}
C.~B. Peel, B.~M. Hochwald, and A.~L. Swindlehurst, ``A vector-perturbation
  technique for near-capacity multiantenna multiuser communication-part {I}:
  channel inversion and regularization,'' \emph{IEEE Transactions on
  Communications}, vol.~53, no.~1, pp. 195--202, Jan 2005.

\bibitem{6375940}
F.~Rusek, D.~Persson, B.~K. Lau, E.~G. Larsson, T.~L. Marzetta, O.~Edfors, and
  F.~Tufvesson, ``Scaling up {MIMO}: Opportunities and challenges with very
  large arrays,'' \emph{IEEE Signal Processing Magazine}, vol.~30, no.~1, pp.
  40--60, Jan 2013.

\bibitem{TulinoMatrix}
A.~M. Tulino and S.~Verd\'{u}, ``Random matrix theory and wireless
  communications,'' \emph{Now Publishers Inc.}, 2004.

\bibitem{Brand200620}
M.~Brand, ``Fast low-rank modifications of the thin singular value
  decomposition,'' \emph{Linear Algebra and its Applications}, vol. 415, no.~1,
  pp. 20--30, 2006.

\bibitem{gradshteyn2007}
I.~S. Gradshteyn and I.~M. Ryzhik, ``Table of integrals, series, and
  products,'' \emph{Elsevier Academic Press, Amsterdam}, 2007.

\bibitem{marzettabook2016}
T.~L. Marzetta, E.~G. Larsson, H.~Yang, and H.~Q. Ngo, \emph{Fundamentals of
  Massive {MIMO}}.\hskip 1em plus 0.5em minus 0.4em\relax Cambridge University
  Press, 2016.

\bibitem{6777295}
J.~Choi, D.~J. Love, and P.~Bidigare, ``Downlink training techniques for {FDD}
  massive {MIMO} systems: Open-loop and closed-loop training with memory,''
  \emph{IEEE Journal of Selected Topics in Signal Processing}, vol.~8, no.~5,
  pp. 802--814, Oct 2014.

\end{thebibliography}

\end{document}